\documentclass[preprint,prb,showpacs,amsmath,amssymb,]{revtex4}
\usepackage{graphics}
\usepackage{epsfig}
\usepackage{subfigure}
\usepackage{amsmath}
\begin{document}
\title{Statistics of Measurement of Non-commuting Quantum Variables:\\
 Monitoring and Purification of a qubit }
\author{ Hongduo Wei and  Yuli V. Nazarov }
\affiliation{Kavli Institute of Nanoscience, Delft University of
Technology, 2628 CJ Delft, The Netherlands}
\date{\today}
\begin{abstract}
We address continuous weak linear quantum measurement and argue that
it is best understood in terms of statistics of the outcomes of the
linear detectors measuring a quantum system, for example, a qubit.
We mostly concentrate on a setup consisting of a qubit and three
independent detectors that simultaneously monitor three
non-commuting operator variables, those corresponding to three
pseudo-spin components of the qubit. We address the joint
probability distribution of the detector outcomes and the qubit
variables. When analyzing the distribution in the limit of big
values of the outcomes, we reveal a high degree of correspondence
between the three outcomes and three components of the qubit
pseudo-spin after the measurement. This enables a high-fidelity
monitoring of all three components. We discuss the relation between
the monitoring described and the algorithms of quantum information
theory that use the results of the partial measurement.

We develop a proper formalism to evaluate the statistics of
continuous weak linear measurement. The formalism is based on
Feynman-Vernon approach, roots in the theory of full counting
statistics, and boils down to a Bloch-Redfield equation augmented
with counting fields.
\end{abstract}
 \maketitle
 \section{introduction}
The theory of quantum measurement, being a foundation of quantum
physics, is  attracting more and more attention
\cite{quantum-measure}.
Intrinsic paradoxes  {\cite{Leggett}} are definitely a main reason for studying
quantum measurements. More motivation comes
from the practical needs to understand the
real solid-state based devices \cite{solid-state,qubit-review}
developed for quantum computing \cite{quantum-computing}.
Measurements in solid-state setups may provide access to extra
variables that facilitate the read out of the quantum information stored
in the elementary
two-level quantum systems (qubits). The concept of continuous weak
linear measurement (CWLM), where the interaction between the
detector and the measured system is explicit and sufficiently weak,
has been recently elaborated in context of the solid state quantum
computing \cite{ Korotkov1, Averin, Clerk, Jordan, Korotkov}.
CWLM provides a universal description of the measurement process and is
based on general linear response theory
\cite{point}. It applies to a large class of linear
detectors: From common amplifiers to more exotic on-chip detectors
such as quantum point contact \cite{QPC}, superconducting SET
transistors \cite{SET}, generic mesoscopic conductors \cite{MES},
fluxons in a Josephson transmission line to measure a
flux qubit \cite{ARS,ARD}.

It is an important feature of CWLM that the (quantum) information is
transferred from a quantum system  being measured --- a qubit --- to
other degrees of freedom: those of the detector. The outcome of the
measurement is thus represented by the detector degrees of freedom
rather than those of the qubit. We will address both the statistics
of the outcomes and joint statistics of the outcomes and the qubit
degrees of freedom.

We stress the difference between the detector outcomes
and the outcomes of a projective measurement of a qubit.
In distinction from the result of a projective measurement,
the detector outcome is not discrete, since the detector output
(for instance, voltage or current) is a continuous variable.
The outcomes do not even have to correlate with the state of the qubit
if the detector is uncoupled. Further, the detector variables are subject to noise
not related to the qubit. Owing to the feedback of the detector at the qubit,
this noise affects the qubit too.

In comparison with the text-book projective measurement that
instantly provides a result and projects the system onto the state
corresponding to the result, the CWLM takes time both to accumulate
the information and to distort the qubit. The time $\tau_{m}$
required to obtain a sufficiently accurate measurement result is
called "measurement time" and is a characteristic of a CWLM setup.
It is not a duration of an individual measurement in this setup: the
latter may vary. The distortion is due to the inevitable  back
action of the detector and is characterized by the dephasing rate
$\Gamma_d$. It has been shown \cite{Korotkov1,Averin,Clerk} that for
an optimized
--- quantum limited --- detector $\tau_{m}\Gamma_d = 1/2$ while the
"measurement time" $\tau_m$ greatly exceeds $1/2\Gamma_d$ for less
optimal detectors.

In the context of quantum information theory, CWLM may be understood
as an interaction of the qubit with infinitely many ancillary qubits
representing the detector degrees of freedom. Each ancilla is
brought to weakly interact with the qubit for a short time and is
subsequently measured. Owing to the interaction, the quantum state
of the ancillae is entangled with the state of the qubit. The
detector output is proportional to the sum of the measurement
results of a large set of ancillae.  This allows to transfer quantum
information from the qubit to the detector without formal projective
measurement of the qubit. Therefore the peculiarities of the CWLM
can be understood in the framework of a projective measurement,
although a more complicated one involving the detector degrees of
freedom. The CWLM can be thus seen as a build-up of an entanglement
between the qubit and the detector. An outcome of an individual
CWLM is the detector output accumulated during the time interval of
a certain duration $\tau_d$. Any CWLM can be described as a
generalized quantum measurement, that involves qubit and detector
degrees of freedom.

The outcome randomly varies from measurement to measurement. We
argue here that studying statistics of the measurement outcomes of a
CWLM is the best way to understand and characterize such a
measurement.  This is especially important for the simultaneous
measurement of non-commuting variables (say, $A$ and $B$) we
concentrate on in this work. In this case, the text-book projective
measurement can not help to predict the statistics of the results:
it would depend on the order of measurements of $A$ and $B$. This
property of the measurements in non-commuting bases enables most
quantum cryptography \cite{crypt} algorithms and has been
extensively elucidated in  Ref. \cite{Busch}.

One can straightforwardly realize in experiment a
CWLM of a quantum system where $A$ and $B$ are measured simultaneously.
 If $A$ and $B$ commute,
the statistics of the outcomes of sufficiently long CWLM corresponds
to the predictions of projective measurement scheme (see Sec. III).
The projective measurement scheme loses its predictive power if $A$
and $B$ do not commute. The reason is that the order of measurement
of $A$ and $B$ is not determined in the course of a continuous
measurement. The statistics of CWLM outputs thus can not be
straightforwardly conjectured and has to be evaluated from the
quantum mechanical treatment of the whole system consisting of the
qubit and the detectors.

In a sharp contrast to the case of commuting variables,
the most probable outcome of a sufficiently
long CWLM of non-commuting variables does not depend
on the qubit state. Therefore it provides no information about the qubit.
The information is however hidden in the statistics of random outcomes.
Recently, the
simultaneous acquisition of two non-communing observables was
investigated in the framework of CWLM {\cite{Jordan}}, and  the
correlation of the random output of two detectors was found to be informative.
Not only noise, but the whole full counting statistics (FCS) of the
non-commuting measurements has been recently addressed for an
example of many spins traversing the detectors \cite{Lorenzo}.

The structure of the article is as follows. We develop the necessary
formalism in Sec. II. Our approach stems from the FCS theory of
electron transfers \cite{Levitov} in the extended Keldysh formalism
\cite{Nazarov}, which has been recently discussed
\cite{Makhlin-Lesovik} in the context of the quantum measurement. At
first step, we obtain a Feynman-Vernon action to describe the
fluctuations of the input and output variables of the detector(s).
In the relevant limit, the action is local in time. So at the second
step we reduce the path integral to the solution of a differential
equation that appears to be a Bloch-Redfield equation augmented with
the counting field. In Sec. III we exemplify the formalism
addressing a relatively simple case of quantum non-demolition (QND)
measurement \cite{QND}. We evaluate the distribution of the outcomes
for a single detector and
 understand  the statistics of  a recently proposed  quantum un-demolition
 measurement\cite{Korotkov}.
The main results are presented in Sec. IV where we discuss
statistics of measurement of non-commuting variables for the case of
three independent detectors measuring the three components of the
qubit pseudo-spin. We find the statistical correspondence between
the three outcomes and three wavefunction components after the
measurement. The correspondence is characterized by a fidelity that
generally increases with the magnitude of the outcomes reaching the
ideal value $1$ in the limit of large magnitudes. Since very large
outcomes are statistically rare and require long waiting times, this
result could be of a purely theoretical value. To prove the
opposite, we have evaluated the fidelity at moderate magnitudes of
outcomes and measurement durations $\tau_d$ and we were able to
demonstrate the fidelity of $0.95$ for $\tau_d \simeq 7 \tau_{m}$.
We term this "quantum monitoring". Ideally, the result of the
quantum monitoring is a pure state of the qubit and three numbers
(detector outputs) giving the polarization of the state. The same
result can be also achieved by preparing the qubit state of the
known polarization, for instance, by a projective measurement along
a certain axis. The difference is that in the case of preparation
the polarization axis is known to the observer in advance, while in
the case of monitoring it is not so: both the three numbers and the
state emerge from dynamics of the quantum system that encompasses
the qubit and the detectors.

We discuss the relation between the quantum monitoring proposed and
the quantum algorithms that use the results of partial measurements
that we summarize in Sec. V.  We evaluate the detector action in the
Appendix A. We prove in the Appendix B that our approach correspond
to a Lindblad scheme for a system consisting of the detectors and
the qubit.

\section{Method}

We start the outline of the formalism with
 a simplest setup where a single detector measures
 a single component of the qubit pseudo-spin.
 In this case, the Hamiltonian reads as
follows:
\begin{subequations}
\begin{eqnarray}
  H &=& H_q+H_{int}+H_{d} \,,\label{Hamiltonian}\\
  H_{q} &=& \sum\limits_{i=1}^{3} H_i \hat{\sigma}_i; \;
  H_{int} = \hat{\sigma}_3 \hat{Q} \label{Hamiltonian-int} \,.
 \end{eqnarray}
\end{subequations}
Here, $H_q$  is the Hamiltonian of the qubit generally given
by a linear combination of three Pauli
matrices $\hat{\sigma}_{i}$ ($i=1$, $2$, $3$) corresponding to three
components of the qubit pseudo-spin.
 $H_{int}$ gives the coupling between the detector and the third component of
the pseudo-spin of the qubit, $\hat{Q}$ being the detector {\it
input} variable.  $H_{d}$ stands for the Hamiltonian of the detector.
Since we assume linear dynamics of the detector variables,
a general form of this Hamiltonian is that of a boson bath,
$$
H_{d} = \sum_k \hbar \omega_k \hat{b}_k^{\dagger} \hat{b}_k.
$$
This encompasses infinitely many boson degrees of freedom labeled by
$k$, $\hat{b}_k$ being the corresponding annihilation operators. The
output of the detector is given by the {\it output} variable
$\hat{V}$. An arbitrary linear dynamics is reproduced if both
variables $\hat{Q}$ and $\hat{V}$ are linear combinations of the
boson creation/annihilation operators,
\begin{eqnarray}
\hat{Q} &=& \sum_k \left( Q_k \hat{b}^\dagger_k + Q^*_k \hat{b}_k \right) \,,\\
\hat{V} &=& \sum_k \left( V_k \hat{b}^\dagger_k + V^*_k \hat{b}_k
\right).
\end{eqnarray}
This is in the spirit of Caldeira-Leggett approach\cite{Caldeira}.
In contrast to the work \cite{Caldeira} we do not assume thermal
equilibrium in the boson bath. In fact, this assumption would be
wrong for most practical detectors since a signal amplification can
not take place in the state of thermal equilibrium. The only
requirement we impose is that Wick's  theorem holds for the boson
operators involved. This guarantees the linear dynamics of the
detector variables.  Besides, this conveniently allows us {\it not}
to specify the coefficients $Q_k, V_k$. All information about the
coefficients and the non-equilibrium boson distribution is
incorporated into the two-point correlators of the variables
explicitly given below (Eqs. 7-8). By virtue of Wick's theorem, the
averages of all possible products of the detector variables can be
expressed in terms of these two-point correlators.

We are interested in the statistics of the detector output variable
$\hat{V}$. We note that this variable is distinct from those of the
qubit and in principle even does not have to correlate with the
qubit state (e.g. if $\hat{Q}=0$). However, since the detector is
supposed to measure the qubit, there must be a (high) degree of
correspondence between the detector output and the qubit state. This
sets the goal of our calculation: to access the joint statistics of
$\hat{V}$ and the qubit variables.

To achieve the goal, we introduce a counting field $\chi(t)$ coupled
to the output variable $\hat{V}$ and use a modified Feynman-Vernon
scheme \cite{Feynman-Vernon} where the evolution of the "bra" and
"ket" wave function is governed by different Hamiltonians $H^{-}$
and $H^{+}$: $H^{\pm}=H\pm \hbar\chi(t) \hat{V}/2$. $\pm$
corresponds to two branches of closed time contour respectively
\cite{Keldysh, Rammer-Smith}. This scheme was first employed in the
work \cite{Nazarov-2}. The counting field $\chi(t)$ plays a role of
the parameter in the generating function of the probability
distribution of the detector outcomes $V(t)$. This generating
function is given by:
\begin{equation}\label{action}
Z(\{\chi(t)\}) =\mathrm{Tr}\bigl( \overrightarrow{\mathrm{T}} e^{
\frac{-i}{\hbar} \int dtH^{-}} \hat{\mathrm{R}}(0)
\,\overleftarrow{\mathrm{T}}e^{ \frac{i}{\hbar}\int    dt H^{+}}
\bigr) \,.
\end{equation}
$\mathrm{Tr}(\cdots)$ implying the trace over  both  detector and
qubit variables. Here, $\overrightarrow{
\mathrm{T}}$($\overleftarrow{\mathrm{T}}$) denotes time (reversed)
ordering in evolution exponents and
$\hat{\mathrm{R}}(0)=\hat{\rho}_d (0) \bigotimes \hat{\rho}(0)$ is
the initial density matrix of whole system. It separates into
$\hat{\rho}_d(0)$  and $\hat{\rho}(0)$, the initial density matrix
of the detector  and qubit respectively. This implies that the
detector and the qubit do not interact before the initial time
moment $t=0$.

Next, we employ the path integral
representation for the probability-generating function
\cite{Nazarov-2}. According to Feynman and Vernon
\cite{Feynman-Vernon} (see also \cite{Kindermann}):
\begin{eqnarray}
Z(\{\chi(t)\})&=& \int\mathcal{D}{\bar X}^{+} \mathcal{D} {\bar
X}^{-} e^{A_d} Z_{Iq}(Q^{-},Q^{+})\,, \label{Z}
\end{eqnarray}
here $\bar{X}^{\pm}(t)$ are two-dimensional vectors of the detector
variables $\bar{X}^{\pm}(t)=(Q^{\pm}(t), V^{\pm}(t))^T$,
$\mathcal{D}{\bar X}^{\pm}\equiv \prod\limits_t dQ^{\pm}(t)d
V^{\pm}(t) $. $Q^{\pm}(t),V^{\pm}(t)$ are the path-integral
variables.
 $Z_{Iq}$ is called the "influence functional" and is
given by
\begin{eqnarray}
 Z_{Iq}( Q^{-}, Q^{+}) &=&\mathrm{Tr}_{\rm q}
   \bigl(\overrightarrow{\mathrm{T}} e^{-\frac{i}{\hbar}\int dt
 \,(H_q + \hat{\sigma}_3 Q^{+}(t)) }\hat{\mathrm{\rho}}(0)\nonumber \\
   & & \times \overleftarrow{\mathrm{T}} e^{\frac{i}{\hbar}\int dt \,(H_q +
\hat{\sigma}_3 Q^{-}(t))
   }
   \bigr)\,,
\end{eqnarray}
where $\mathrm{Tr}_q$ means the trace over qubit space. The action
$A_d$ in Eq. (\ref{Z}) is bilinear in $\bar{X}^\pm$ and $\chi$ to
conform to linear dynamics of the detector and will be specified
below. The advantage of this representation is that the dynamics of
infinitely many detector degrees of freedom have been reduced to the
dynamics of only two relevant fields: $\hat{Q}$ and $\hat{V}$. The
influence functional describes a non-linear response of the qubit on
the fields.

Let us turn to a specific model of linear dynamics of the detector.
Following common assumptions about CWLM, \cite{Averin,Clerk} we
assume instant detector responses and white (frequency-independent)
noises. Under these assumptions, a detector is characterized by
seven independent parameters: four response functions and three
noises. It is convenient to use an index $i$ taking values $1$ and
$2$ for input and output variables respectively. With this index, we
present $4$ response functions $a_{ij}$ as a single $2 \times 2$
matrix.  The noises $S_{ij}$ form a similar matrix. By virtue of
Kubo formula, the response functions are expressed in terms of
expectation values of the operator commutators
\begin{subequations}
\label{responses}
\begin{eqnarray}
-\frac{i}{\hbar}\langle[\hat{Q}(t),\hat{Q}(t')] \rangle&=&a_{11}\delta(t-t'-0^{+})\,,\\
-\frac{i}{\hbar}\langle[\hat{Q}(t),\hat{V}(t')] \rangle&=&a_{12}\delta(t-t'-0^{+})\,,\\
-\frac{i}{\hbar}\langle[\hat{V}(t),\hat{Q}(t')] \rangle&=&a_{21}\delta(t-t'-0^{+})\,,\\
-\frac{i}{\hbar}\langle[\hat{V}(t),\hat{V}(t')]\rangle&=&a_{22}\delta(t-t'-0^{+})\,,
\end{eqnarray}
\end{subequations}
where an infinitesimal  small positive number $0^{+}$  in the
$\delta$-function  represents small but finite response time.

The noises correspond to the expectation values of symmetrized
operator products,
\begin{subequations}
\label{noises}
\begin{eqnarray}
\langle\langle \frac{\hat{Q}(t) \hat{Q}(t')+\hat{Q}(t') \hat{Q}(t)
}{2}\rangle\rangle &=&S_{11}\delta(t-t') \,,\\
\langle\langle \frac{\hat{V}(t) \hat{V}(t')+\hat{V}(t') \hat{V}(t)
}{2}\rangle\rangle &=&S_{22}\delta(t-t') \,,\\
\langle\langle \frac{\hat{Q}(t) \hat{V}(t')+\hat{V}(t') \hat{Q}(t)
}{2}\rangle\rangle &=&S_{12}\delta(t-t') \,,\\
\langle\langle \frac{\hat{V}(t) \hat{Q}(t')+\hat{Q}(t') \hat{V}(t)
}{2}\rangle\rangle &=&S_{21}\delta(t-t') \,,
\end{eqnarray}
\end{subequations}
here, $\langle\langle \hat{A}\hat{B}\rangle\rangle\equiv \langle
(\hat{A}-\langle \hat{A}\rangle )(\hat{B}-\langle \hat{B}
\rangle)\rangle$ for  any operators $\hat{A}$ and $\hat{B}$.

Let us discuss the physical meaning of the parameters involved.
$S_{11}$ is the noise of the input variable
responsible for the back action of the detector and decoherence of
the qubit; $S_{22}$ is the output noise that prevents a fast measurement
of the detector outcome. The cross-term $S_{12}=S_{21}$ presents
the correlation of these two noises.
The response function $a_{21}$ determines the
detector gain: The proportionality coefficient
between the detector output and the third component
of the qubit pseudo-spin, $\langle \hat{V} \rangle
= a_{21} \langle \hat{\sigma}_3\rangle$.
Other response functions
$a_{12},a_{22},a_{11}$ are respectively related to reverse gain,
output and input impedances of the detector and are not of immediate
interest for us.
The detector is characterized with the dephasing
rate $\Gamma_d=2 S_{11}/\hbar^2$ and the "measurement time" $\tau_m=
S_{22}/a_{21}^2$ \cite{Makhlin-2, Averin}. The Cauchy-Schwartz
inequality
$$S_{11}S_{22}-S_{12}^2\geq \frac{\hbar^2}{4}
(a_{21}-a_{12})^2$$ imposes an important restrictions on the
possible values of the parameters \cite{Averin, Clerk}. Following
the common assumption, we assume  that the reverse gain $a_{12}$ is
much less than the direct  gain $a_{21}$: $a_{21}\gg a_{12}$. This
condition is commonly required from a good amplifier. Under these
assumptions, $\tau_m \Gamma_d \ge 1/2$: one cannot measure a qubit
without dephasing it.

The  action $A_d$
corresponding to the model reads
\begin{eqnarray}\label{newaction1}
A_d = \int dt [ -\frac{1}{2}\bar{x}^{T}(t)
(\check{a}^{-1})^{T}\check{S}\check{a}^{-1}\bar{x}(t)
+i\bar{X}^{T}(t)\check{a}^{-1}\bar{x}(t)+i
\bar{\chi}^{T}(t)\bar{X}(t) ]\,,
\end{eqnarray}
the derivation  is outlined in the appendix A.
Here we switch to the "quantum" $\bar{x}$
 and "classical" $\bar{X}$
variables defined as follows: ${\bar x}=({\bar X}^{+}-{\bar
X}^{-})/\hbar$, ${\bar X}=({\bar X}^{+}+{\bar X}^{-})/2$.
$\bar{\chi}=(0,\,\chi)^{T}$. The $2\times 2$ matrices $\check{a}$
and $\check{S}$  are respectively:
\begin{subequations}
\begin{eqnarray}
\check{a}&=& \left(
               \begin{array}{cc}
                 a_{11} & a_{12}\\
                 a_{21} & a_{22}\\
               \end{array}
             \right)\,,\\
\check{S}&=& \left(
               \begin{array}{cc}
                 S_{11} & S_{12} \\
                 S_{21} & S_{22} \\
               \end{array}
             \right)\,.
\end{eqnarray}
\end{subequations}

It is important for further advance that the action
(\ref{newaction1}) is local in time. This allows for reducing the
path integral to a differential equation. The procedure is
completely similar to the standard reduction of the corresponding
path integrals to either Schr\"{o}dinger or Fokker-Planck
equations\cite{Kleinert}. One slices time axis into intervals
$(t_{k},\, t_{k+1})$ ($t_k= k \Delta t$. $k$ is an integer), and
takes the path integral in (\ref{Z}) without tracing over the qubit
indexes slice by slice. The result of the integration at $t_k$ is a
matrix in qubit indexes, $\hat\rho(t_k)$. Integrating over $x,X$ in
the next slice, one finds a linear relation between
$\hat{\rho}_{t_{k+1}}$ and $\hat{\rho}_{t_{k}}$:
\begin{eqnarray}
\hat{\rho}(\chi;t_{k+1})=\int \prod\limits_{ t_{k}<t<t_{k+1}}
\mathcal{D} \bar{x}(t) \mathcal{D} {\bar X}(t)  e^{A_d} \hat{S}_{+}
\hat{\rho}(\chi; t_{k})\hat{S}_{-};\\
\hat{S}_{\mp} = \exp\left(\pm \frac{i}{\hbar} \left( H_q \Delta t
+\hat{\sigma}_3  \int\limits_{t_{k}}^{t_{k+1}} dt Q^{\mp}(t)
\right)\right)\,.
\end{eqnarray}
Since the slice is thin, the exponents may be expanded,
$$
\hat{S}_{\mp} \approx \hat{1} \mp \frac{i}{\hbar} \left( H_q \Delta
t +\hat{\sigma}_3  \int\limits_{t_{k}}^{t_{k+1}} dt Q^{\mp}(t)
\right) - \frac{1}{2\hbar^2}\int\limits_{t_{k}}^{t_{k+1}} dt  dt'
Q^{\mp}(t) Q^{\mp}(t')+ \dots
$$
and the integration is reduced to evaluation of
the averages and the correlators of the fields $Q^{\pm}$ with the action
$A_d$.
Collecting terms $\propto \Delta t$ and taking the limit
$\Delta t \to 0$,
we obtain a differential equation for $\rho(t)$ that resembles a familiar
Bloch-Redfield equation but essentially depends
on the counting field $\chi$ \cite{Romito}.
The resulting equation for $\rho(\chi;t)$ reads:
\begin{eqnarray}
      \frac{\partial{\hat{\rho}}}{\partial t} &=&
      -\frac{i}{\hbar} [H_q,\hat{\rho}
      ]-\frac{\chi^2(t)}{2}S_{22}\hat{\rho}
+ \frac{i a_{21}\chi(t)}{2}(\hat{\rho} \hat{\sigma}_3 +
\hat{\sigma}_3
\hat{\rho}) \nonumber \\
        & &-\frac{S_{12}}{\hbar}\chi(t)(\hat{\rho} \hat{\sigma}_3 - \hat{\sigma}_3 \hat{\rho})
        -\frac{S_{11}}{\hbar^2}(\hat{\rho}-  \hat{\sigma}_3
        \hat{\rho}\hat{\sigma}_3
        ) \,.
\label{master-eq}
\end{eqnarray}

How to apply the equation? Let us consider a single measurement first.
Let $\tau_d$ be the duration of the measurement. We collect the detector
output during the time interval $(0,\tau_d)$ and normalize it by $\tau_d$
\begin{eqnarray}
V_o=\frac{1}{\tau_d}\int_{0}^{\tau_d} V(t)\, dt\,.
\end{eqnarray}
To get the statistics of $V_o$, we should assume that $\chi$ is a
constant in the interval $(0,\tau_d)$. Indeed, expanding of the
generating function in terms of $\chi$ gives the averages of
products of $V_o$. Let us suppose that the initial density matrix of
the qubit is $\hat\rho(0)$. We solve the Eq. (\ref{master-eq}) with
the initial conditions $\hat\rho(\chi,0)=\hat\rho(0)$. The output is
a $\chi$-dependent matrix $\hat\rho(\chi) \equiv
\hat\rho(\chi,t=\tau_d)$ after the measurement.

We stress that $\rho(\chi)$ appearing in the
equation is {\it not} the reduced qubit density
matrix. It is a more interesting and complicated quantity that reflects
the joint probability distribution of the qubit pseudo-spin components
after the measurement and the detector outcome collected.
To see this, let us define the reduced density matrix of the qubit and the
outcome $V_o$, $\hat{R}(V_o,V'_o)$. It is a matrix in qubit indices
and in the outcome values $V_o,V'_o$ (see Appendix B for the details). Its diagonal
elements give the statistics in question.
The reduced qubit density matrix (with no regard for the value of the outcome)
is given by
$$
\hat{\rho}(t) = \int d V_o \hat{R}(V_o,V_o),
$$
the probability distribution of the outcomes (with no regard for the qubit state)
reads
$$
P(V_o) = {\rm Tr}_q \hat{R}(V_o,V_o),
$$
and the joint statistics is expressed in terms of the qubit density matrix
{\it conditioned} to a certain value of the output,
$$
\hat{\rho}(V_o) = \hat{R}(V_o,V_o)/P(V_o).
$$
The quantity in use, $\hat \rho(\chi)$, is related  to the diagonal
elements of thus introduced density matrix $\hat{R}$ by means of
Fourier transform \cite{Nazarov-2},
\begin{eqnarray}
\hat{R}(V_o,V_o)= \frac{t}{2\pi}\int d \chi\, \hat{\rho}(\chi) e^{-i
\chi V_o t}\label{Fourier}
\end{eqnarray}
presenting a generating function for the quasi-distribution
$\hat{R}(V_o,V_o)$. Comparing this with the above definitions, we
find convenient relations
$$
\hat \rho(t)=\hat\rho(\chi=0),\ P(V_0) = t\int \frac{d \chi}{2\pi}\,
e^{-i \chi V_o t} {\rm Tr}_q \hat{\rho}(\chi),$$
$$\hat{\rho}(V_o)=\frac{\int d \chi  e^{-i\chi V_o
t}\hat{\rho}(\chi)}{\int d \chi\, \mathrm{Tr_{q}\hat{\rho}(\chi)
e^{-i \chi V_o t}}}.$$ It is the main technical advantage of our
work that Eq. \ref{master-eq} is similar in form to an elementary
Bloch-Redfield equation for the qubit density matrix and not much
complicated than that one. However, in its augmented form it solves
a much more challenging task of finding the joint probability
distribution of the detector outcome and the qubit state.

It is important to note that  Eq. (\ref{master-eq}), as well as Eq.
(\ref{master-eq2}) for multi-detector setup, complies with a
Lindblad scheme \cite{Lindblad,Presilla}. We will show this
explicitly in the appendix B. This guarantees the positivity of  the
"big" density matrix $\hat{R}(V_o,V_o)$.

The locality in time is a relevant but strong assumption which in
fact corresponds to a  {\it classical} detector (indeed, the action
(\ref{newaction1}) does not contain any $\hbar$).  This is why we do
not have to worry about possible quantum uncertainties of the
detector output that could complicate the interpretation of the
statistics  \cite{Nazarov-2}.

The scheme can be easily extended to many repetitive (that is, being
constantly repeated) measurements to comply with the concept of
CWLM. Let us consider (infinitely) many subsequent measurements. For
$i$-th measurement, the detector output is collected during the time
interval $(t_i,t_{i+1})$. This gives a series of outcomes
$V^{(i)}_o$. To describe the joint statistics, one solves Eq.
(\ref{master-eq}) with a piece-wise constant $\chi(t)$, $\chi(t) =
\chi_i$ in the interval $(t_i,t_{i+1})$. The solution of the
equation at the time moment $t_{M+1}$ depends on $M$ counting
fields: $\hat\rho = \hat\rho(\chi_1,\dots,\chi_M)$. The Fourier
transform with respect to all $\chi_i$ defines the qubit density
matrix {\it conditioned} on the outcomes $V^{(i)}_{o}$ of all $M$
preceding measurements. We illustrate two subsequent measurements in
the next Section.

Importantly, the scheme described can also be easily
extended to more qubits and/or detectors: One just adds extra
(counting) fields for detectors and extra Pauli matrices for qubits.
The case of interest for us is the simultaneous CWLM of three
pseudo-spin projections of a qubit.
The coupling term becomes
\begin{equation}
\label{three-detectors}
 H_{int}=\hat{\sigma}_1 \hat{Q}_1 +\hat{\sigma}_2 \hat{Q}_2+\hat{\sigma}_3 \hat{Q}_3 \,.
\end{equation}
 $\hat{Q}_k$ ($k=1$, $2$, $3$)
being the input fields of the
three detectors. Three counting fields $\chi_k$  are coupled to
the corresponding output variables $\hat{V}_{k}$  of the three detectors.

While it is straightforward to write down the equation for general
situation, we employ a specific model at this point.
Namely, we assume for simplicity that the detectors are
identical and independent.  "Identical" implies that
the noises and response functions of all
three detectors are the same.
"Independent" implies that no response function relates
inputs/outputs of two different detectors, neither the
noises correlate. Each detector is described by the action
in corresponding variables. Under these assumptions, the
setup is conveniently $SU(2)$ covariant.

 The resulting $\chi$-augmented Bloch-Redfield equation
 reads
\begin{eqnarray}
      \frac{\partial{\hat{\rho}}}{\partial t} &=&
      -\frac{i}{\hbar} [H_q,\hat{\rho}
      ] -(\sum_{k=1}^{3}\frac{\chi_k^2}{2})S_{22}\hat{\rho}
+ \frac{i a_{21}}{2}\sum_{k=1}^{3}\chi_k
[\hat{\sigma}_k,  \hat{\rho}]_{+}\nonumber \\
        & &+\frac{S_{12}}{\hbar}\sum_{k=1}^{3} \chi_k[ \hat{\sigma}_k,
        \hat{\rho}]
        -\frac{S_{11}}{\hbar^2}(3\hat{\rho}-\sum_{k=1}^{3}\hat{\sigma}_k
    \hat{\rho}\hat{\sigma}_k)\,.
\label{master-eq2}
\end{eqnarray}
 Comparing
this with Eq. \ref{master-eq}, we see that each detector
contributes a term to the equation. Each term comes with the corresponding
counting field and $\sigma$-matrix.

\section{Statistics of QND measurement}

Before turning to the measurements of non-commutative variables, let
us first illustrate the formalism with a single-detector setup. Only
a single component $\hat{\sigma}_3$ will be measured. We will be
interested in a quantum QND setup where successive measurements are
performed. Such QND  measurements have been recently realized for
superconducting qubits \cite{QNM-hans}. To satisfy the
non-demolishing condition \cite{QND}, we set $\hat{H}_q = \epsilon
\hat{\sigma}_3$ in (\ref{Hamiltonian}) so that $H_q$ and $H_{int}$
commute. In this case, $H_q$ can be canceled by transformation to
the rotating frame, $\hat{\rho} \to e^{i\hat{H}_qt/\hbar}\hat{\rho}
e^{-i\hat{H}_qt/\hbar}$ and will be disregarded from now on.

Let us
perform two measurements that immediately follow each other. During
the first measurement of duration $t_1$, the detector output is
collected in the time interval $(0,t_1)$ so the measurement outcome
is $V_1=\int_0^{t_1} dt V(t)/t_1$. Similarly, for the second
measurement $V_2=\int_{t_1}^{t_1+t_2} dt V(t)/t_2$. The statistics
of the two outcomes is computed from (\ref{master-eq})
by setting $\chi(t)$ to a piece-wise constant  $\chi(t)
=\chi_1(\chi_2)$ during the first(second) time interval and
$\chi(t)=0$ otherwise. To solve the equation,
we parameterize $\hat{\rho}(\chi ;t)$ as
follows:
\begin{eqnarray}
\hat{\rho}(\chi; t)&=&\frac{\hat{1}+\hat{\sigma}_3}{2}\rho_{+}(\chi;
t)+\frac{\hat{1}-\hat{\sigma}_{3}}{2}\rho_{-}(\chi; t) \nonumber\\
& &+\hat{\sigma}_{1}\rho_{1}(\chi; t)+\hat{\sigma}_{2}\rho_{2}(\chi;
t)\,, \label{rho-parameter}
\end{eqnarray}
where $\hat{1}$ is the unit $2\times 2$ matrix in the qubit space,
$\rho_{\pm}$ give diagonal elements of the matrix and
$\rho_{1,2}$ give the non-diagonal ones.
 Eqs. (\ref{rho-parameter}) and (\ref{master-eq}) yield
 two pairs
 of separating
 equations:
\begin{subequations}
\begin{eqnarray}
 \frac{\partial \rho_{+}}{\partial t}&=& i\chi a_{21} \rho_{+}-\frac{\chi^2}{2}S_{22}\rho_{+} \,,
  \label{rho-a} \\
\frac{\partial \rho_{-}}{\partial t} &=& -i\chi
a_{21}\rho_{-}-\frac{\chi^2}{2}S_{22}\rho_{-} \,,
\end{eqnarray}
\end{subequations}
and
\begin{subequations}
\begin{eqnarray}
\frac{\partial \rho_{1}}{\partial t}&=& -\frac{2i S_{21}\chi}
{\hbar} \rho_{2}-(\Gamma_d +\frac{\chi^2}{2}S_{22})\rho_{1}\,, \label{rho-1}\\
\frac{\partial \rho_{2}}{\partial t}&=& \frac{2i S_{21}\chi} {\hbar}
\rho_{1}-(\Gamma_d +\frac{\chi^2}{2}S_{22})\rho_{2}\,. \label{rho-2}
\end{eqnarray}
\end{subequations}

We first assume that the initial density matrix of the qubit is diagonal.
We solve  $\rho_{\pm}(\chi;t)$ with initial conditions
$\rho_{\pm}(0)$ at $t=0$. We note that at the initial time $t=0$,
$\hat{\rho}$ does not depend on $\chi$ or $V_o$ and thus we do not
write them explicitly for the initial conditions. Solving the
equations and Fourier-transforming the generating function
yields a very
simple probability distribution of both outcomes:
\begin{equation}
P(v_1,v_2) = \sum\limits_{\pm}
\frac{\sqrt{\tau_1\tau_2}}{2\pi}\rho_{\pm}(0)
e^{-\frac{(v_1\mp1)^2\tau_1}{2}} e^{-\frac{(v_2\mp1)^2\tau_2}{2}}\,.
\end{equation}
To keep it simple, we have switched here to the dimensionless
durations $\tau = \tau_d/\tau_m$, and outcomes $v = V_o/a_{21}$. The
result is in fact classical: it does not depend on the dephasing
rate. In allows for an elementary interpretation: initially, the
qubit appears to be  either in the state $+$ or $-$, with
probabilities $\rho_{+}(0)$ and $\rho_{-}(0)$ respectively. The
state persists during the measurements. The outcome of each
measurement is distributed normally around $\pm 1$ with the standard
deviation $\sqrt{\tau_{1,2}}$ set by the duration of the
measurement. We note that the persistence of a state is specific for
QND measurements, and, as we see in the next Section, does not apply
to the CWLM of non-commuting variables.

In Fig. 1, the solid lines show  the distribution of outcome $P$
versus $v$  for  two different durations $\tau_1$=2 (long, lower
curve) and $\tau_1=0.3$ (short, upper curve). Two  obvious peaks
located at $v=\pm 1$ for the line $\tau_1=2$ are due to the two
states of the qubit that can be distinguished  in the course of the
long measurement. For the short measurement $\tau_1=0.3$, the
detector can not resolve the difference between two eigenstates of
the qubit. Thus we see a single peak at $v=0$ broadened by the noise
of the output variable. The dotted lines show the probability
distribution $P$ of the outcome of the  second measurement of the
same duration {\it under condition} that the outcome of the first
measurement $v_1=-1$. For long measurement ($\tau_{2}=2$), the
probability distribution is concentrated near $v=-1$. For the short
measurement ($\tau_2=0.3$), the distribution is similar to that of
the first measurement, its average being close to $v=0$. This makes
a comprehensive illustration  of the fact that the sufficiently long
measurements are repeatable, that is, the result of the second
measurement is close to the result of the second one. This is not
the case for the short measurement.

\begin{figure}[htb]
\begin{tabular}{cc}
 \resizebox{84mm}{!}{\includegraphics{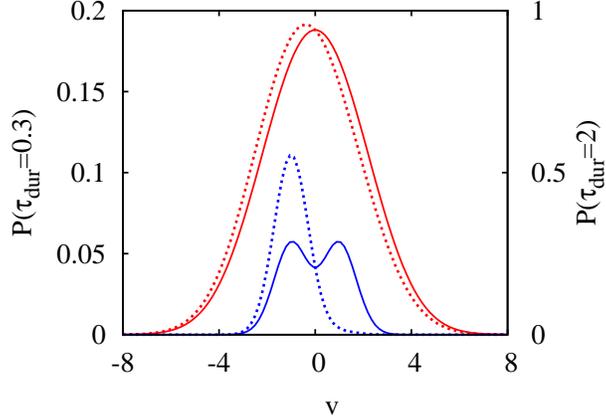}}
   \end{tabular}
 \caption{
 (color online) Quantum non-demolition setup:
 Two successive measurements. In each pair of the curves,
 the solid one gives the distribution of outcome of the first
 measurement while the dashed one gives the distribution for
 the second measurement {\it provided} the first measurement
 gave $v_1=-1$.
 Lower (upper) pair of curves corresponds to long, $\tau_{1,2}=2$
 (short, $\tau_{1,2}=0.3$)
 duration of measurement. The long measurement is thus repeatable,
 the short one is not.
 }
\end{figure}

To illustrate the quantum aspect, let us set the initial density
matrix to correspond to a pure state with
the wave function that is an equal superposition of
the base states $\pm$:
$\hat\rho(0)=\hat{\sigma}_1$.
We are interested in the average value of the
corresponding pseudo-spin projection $\hat{\sigma}_1$
after the measurement
{\it provided} the measurement gives the outcome $v$.
We evaluate this average if we know
the qubit density matrix $\hat{\rho}(v)$
{\it conditioned} on the outcome $v$,
\begin{eqnarray}
\sigma_1(v)\equiv \mathrm{Tr}(\hat{\sigma}_1
{\hat\rho}(v)).
\end{eqnarray}
As discussed in the previous Section, this qubit density
matrix is computed from the normalized Fourier transform
of $\hat{\rho}(\chi)$, the latter is obtained
by solving Eqs. (\ref{rho-1}) and (\ref{rho-2})
on time interval $(0,\tau)$, $\tau$ being duration of
the measurement.

Since Eqs. (\ref{rho-2}) do contain
the dephasing rate, the result will depend on actual depahsing.
The answer reads
\begin{eqnarray}
\sigma_1(v;\tau) &=&\frac{\cos (C_{12}v\tau) }  {\cosh(v\tau)}
e^{-\frac{C}{2}\tau}\,.
\end{eqnarray}
Here, we introduce dimensionless constants $C \equiv
4(S_{11}S_{22}-S_{12}^2)/(\hbar a_{21})^2 -1$ and $C_{12} =2
S_{12}/\hbar a_{21}$. $C>0$ characterizes the quality of
the detector, $C=0$ for a quantum-limited one. $C_{12}$
characterizes the correlations of the noises, for
a quantum-limited detector $C_{12}=0$ as well.

Generally, $\sigma_1(v,\tau)$ quickly decays with increasing $\tau$.
This manifests the dephasing of the superposition by the
measurement. Remarkably enough, for a quantum-limited detector
($C=C_{12}=0$) and for a special value of the measurement outcome
$v=0$ the dephasing is absent. The wave function retains its initial
value for this particular value of output. This fact has been noted
in the work \cite{Korotkov} and termed "quantum un-demolition
measurement". Let us note  that the phase shift between the states
$\pm$, acquired from the detector, $2\int_0^{\tau}\,dt Q(t)/\hbar$,
is zero at this (rather improbable  \cite{Korotkov}) value of the
outcome. We stress that the strict correspondence between the phase
shift and outcome does not hold for a general detector, so that
$\sigma_1(v=0, \tau) = \exp(-C\tau/2)$ decreases with the
measurement duration $\tau$ indicating the non-vanishing dephasing
of the superposition. We plot $\sigma_1(v;\tau=1)$ versus the
detector outcome $v$ for
 the duration of measurement $\tau=1$ in Fig. 2.
The solid line is for the  quantum-limited detector ($C=C_{12}=0$),
 and the dotted line is for the a worse detector
 ($C=C_{12}=1$).
\begin{figure}[htb]
  \begin{tabular}{cc}
 \resizebox{84mm}{!}{\includegraphics{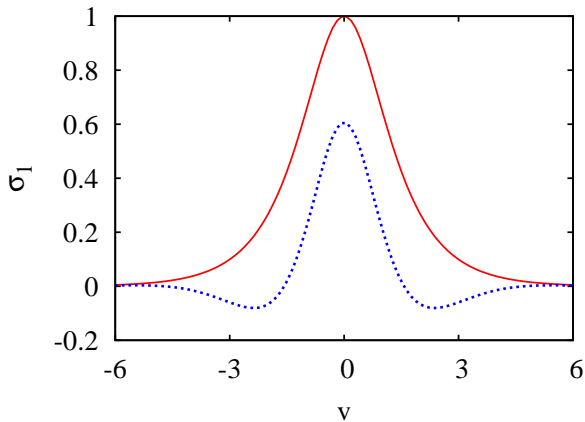}}
   \end{tabular}
 \caption{
 (color online)
 "Un-demolition" measurement.
 The average value of pseudo-spin component
 $\sigma_1(v; \tau)$ {\it conditioned} on
 the detector outcome  $v$ characterizes the dephasing of
 the superposition after a QND
 measurement of duration $\tau$($\tau=1$ for the plots).
 A quantum-limited detector ($C=C_{12}=0$, upper curve)
 allows for the quantum "un-demolition"
 measurement ($\sigma_1=1$ at $v=0$).
 This does not work for a worse detector ($C=C_{12}=1$, lower curve).
 }
\end{figure}

\section{The statistics of the CWLM of
non-commuting variables}

In the previous  Section we provided simple examples to prove the
use of the statistical approach. Thus encouraged, we turn to the
statistics of
the CWLM of non-commuting variables. The interaction Hamiltonian is
now given by Eq. (\ref{three-detectors}).  We do not want
to deal with the qubit Hamiltonian $H_q=\epsilon \sigma_3$
and shall assume that it is removed by transforming to
the rotating frame. The same transform makes $\sigma_{1,2}$
to rotate with angular velocity $\epsilon/\hbar$.
To compensate for this, let us presume that the signal from
$\sigma_{1,2}$ is collected at frequencies $\epsilon/\hbar$ rather
than at zero frequency as the signal from $\sigma_3$ is.
Mathematically, we define the outcomes of the detectors
$1,2$ as
\begin{eqnarray}
\tau_d V_{1,2} = \int_0^{\tau_d}\, dt \bigl[\cos(\frac{\epsilon
t}{\hbar}) V_{1,2}(t) \mp \sin(\frac{\epsilon t}{\hbar})
V_{2,1}(t)\bigr]\,.
\end{eqnarray}
This can be practically realized in a very same way
as it is done in a radio set. One has to mix
a high frequency signal with a reference signal of the same
frequency and detect the low-frequency component of the product.

Let us evaluate $\hat{\rho}(\chi_{1},\chi_{2},\chi_3)$. Without the
term $H_q$, the Eq. (\ref{master-eq2}) is readily solved in a proper
basis in the pseudo-spin space. In this basis, one of the Pauli
matrices is defined as $\hat{\sigma}_{\chi}= (\chi_1 \hat{\sigma}_1
+\chi_2 \hat{\sigma}_2+\chi_3 \hat{\sigma}_3)/\chi_s$,  while two
others, $\hat{\sigma}_{\mu}$ and $\hat{\sigma}_{\nu}$, are chosen to
be orthogonal to it. We have introduced $\chi_s$ as follows:
$\chi_s\equiv\sqrt{\chi_1^2+\chi_2^2+\chi_3^2}$.  We parameterize
$\hat{\rho}$ as follows:
\begin{eqnarray}
  \hat{\rho}=\rho_0\hat{1}+\rho_{\chi}\hat{\sigma}_{\chi}+\rho_{\mu}\hat{\sigma}_{\mu}
  +\rho_{\nu}\hat{\sigma}_{\nu} \,.
\label{pameter-rho2}
\end{eqnarray}
From Eqs. (\ref{master-eq2}) and (\ref{pameter-rho2}) we  obtain two
 equations involving $\rho_0$ and $\rho_{\chi}$:
\begin{subequations}
\begin{eqnarray}
\frac{\partial \rho_0}{\partial t}&=& i a_{21} \chi_s \rho_{\chi}
-\frac{\chi_s^2}{2}S_{22}\rho_{0} \,,
\label{rho-chi-1}\\
\frac{\partial \rho_{\chi}}{\partial t}&=& i a_{21} \chi_s \rho_0
 -(2\Gamma_d+\frac{\chi_s^2}{2}S_{22})\rho_{\chi}
 \,.
\label{rho-chi-2}
\end{eqnarray}
\end{subequations}

We stress that the CWLM we are about to describe is hardly a
measurement of the {\it initial} state of the qubit. In contrast to
QND where the dephasing is limited to $1,2$ components, the input
variables $\hat{Q}_{1,2,3}$ of the detectors randomly rotate the
pseudo-spin in all three directions. The quantum information about
initial state is lost rather quickly: at the time scale of
$1/\Gamma_d$. That is, it is lost before a statistically reliable
measurement result can be accumulated. This motivates us to choose
the {\it unpolarized} density matrix as initial condition for the
state of the qubit before the measurement,
\begin{eqnarray}
\hat\rho(0) = \frac{1}{2}\hat 1\,. \label{rho-0-3d}
\end{eqnarray}
We will see however that despite the memory lost
the CWLM of non-commutative variables can be rather informative.

\begin{figure}[htb]
  \begin{tabular}{cc}
  \resizebox{84mm}{!}{\includegraphics{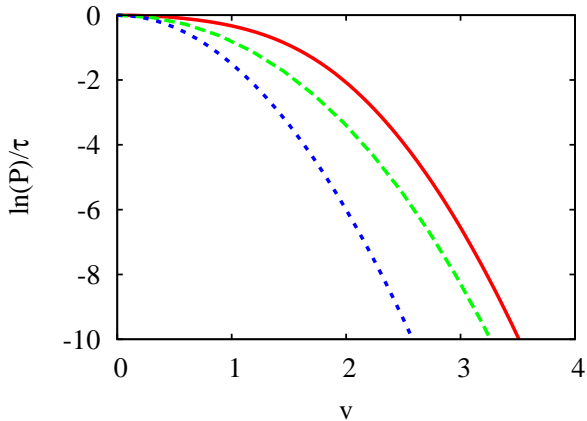}}
   \end{tabular}
\caption{(color online)  Logarithm of the outcome distribution. The
curves from the top to the bottom: quantum-limited detector
($C_d=1/2$), worse detector ($C_d=2$), detector not connected to the
qubit.}
\end{figure}

Let us first discuss the distribution of the detector outputs. We
first solve   Eqs. (\ref{rho-chi-1}) and (\ref{rho-chi-2}) with the
initial condition (\ref{rho-0-3d}), and then recall  Eq. (\ref{Z}).
In the limit of long durations $\tau \gg 1$, the log of the
generating function reads:
\begin{equation}\label{S}
-\log Z =\tau\left( C_d
-\sqrt{C^2_d-\chi_s^2}+\frac{\chi_s^2}{2}\right) \,,
\end{equation}
where $\chi_i$ has been made dimensionless $\chi_i S_{22}/a_{21} \to
\chi_i $ as to give the cumulants of dimensionless outputs $v_i$.
Here, $C_d \equiv \Gamma_d\tau_m  = (C+1+C_{12}^2)/2\geq 1/2$.

The cumulants of the outcomes can be evaluated by taking the
derivatives of the $\log Z$ with respect to $\chi_i$. The presence
of the qubit enhances output noises of each detector by the factor
$1+1/C_d$:
\begin{eqnarray}
 \langle\langle v_i v_j \rangle\rangle &=&- \frac{\partial^2
 \log Z}{ \partial \chi_i \partial
 \chi_j}|_{\chi_{i,j}=0}
 =\frac{1}{\tau}(1+\frac{1}{C_d})\delta_{ij}\,.
\end{eqnarray}
There is no correlation of noises between different detectors. Such
correlation arise for  fourth cumulants
\begin{eqnarray}
 \langle\langle v^2_i
v^2_j\rangle\rangle =-\frac{\partial^{4}\log Z}{\partial \chi_i^2
\partial \chi_j^2}|_{\chi_{i,j}=0}= -\frac{1+2\delta_{ij}}{(C_d\tau)^3}\,.
\end{eqnarray}
The distribution is isotropic in three outputs depending on $v\equiv
\sqrt{v_1^2+v_2^2+v_3^2}$ only and in the limit of long durations we
can calculate it by the saddle-point method, determining an optimal
$\chi^{*}$ corresponding to a given outcome $v$. We obtain
\begin{eqnarray}
 \log P(v)=\log Z(\chi^{*}(v))\,;\,\, \,
 \frac{\partial \log Z}{\partial \chi_s}|_{\chi_s=\chi^{*}}
  = i v\tau
 \,.
\end{eqnarray}
$\chi^{*}$ is purely imaginary. We plot ${\log (P)}/\tau$ vs. $v$
for three different values of $C_d$ in Fig. 3.  The solid line is
for the quantum-limited detector ($C_d=1/2$), the dashed line is for
the worse detector ($C_d=2$). The dotted line is for  the detectors
not connected to the qubit ($C_d=\infty$). So it is a parabola
corresponding to the Gaussian distribution of outcomes in this case.
We see that the distribution is concentrated at zero. Typical values
of outcomes $v\sim 1/\sqrt{\tau}$ and for these typical values the
distribution can be approximated estimated by a Gaussian  one $P\sim
\exp{(-v^2/2 (1+1/C_d))}$. At larger (and thus atypical) values of
outcomes ($v\simeq 1$) the distribution is essentially non-Gaussian.
We see from the plots that the presence of the qubit exponentially
enhances probabilities of such outcomes.

Let us discuss the correlation of
the detector outputs and the pseudo-spin {\it after} such
measurement thus turning to the joint statistics of
the measurement outcomes and the resulting qubit state.

We characterize the correlation with a fidelity $f(v)$, inner
product of the normalized vector of the outcomes and averaged
pseudo-spin at {\it given} outcome $v$:
\begin{eqnarray}
f=\frac{\sum_i\langle \sigma_i \rangle v_i}{v} \,;\,\,\,\, \langle
\sigma_i \rangle=\mathrm{Tr}_q [\hat{\sigma}_i \hat{\rho}(v)]\,,
\end{eqnarray}
where $\hat{\rho}(v)$ is defined in Eq. (\ref{Fourier}) The fidelity
is $1$ if the normalized
 values of the outputs precisely give
all three pseudo-spin components. Analyzing the saddle-point
solution for the $\hat{\rho}(v;\tau)$, we obtain that $f$ does not
depend on $\tau$ in the limit $\tau \gg 1$. Importantly,  at large
values of the outcomes $v\gg 1$ the fidelity reaches the ideal value
$f \approx 1 - C_d/v$. This, quite unexpectedly, enables an
efficient {\it quantum monitoring} of non-commuting variables.

The monitoring procedure is as follows. Starting from some initial
state, one performs a series of repetitive measurements of duration
$\tau$. The three outcomes $v_i$ of each measurement are written
down. For most measurements, the values of the outcomes  are
typical, that is, do not exceed the results of such measurements
correspond to low fidelity and are therefore discarded. One
specifically waits for a measurement that gives sufficiently big
values of outcomes. To decide if the outcomes are sufficiently big,
one estimates the fidelity of each measurement given the values of
outcomes and the relation $f(v)$. If one wants to achieve the
desired  accuracy $a_{des}$, one thus waits for the outcomes
satisfying $f(v)>1-a_{des}$. The big values of outcomes guarantee
that the fidelity $f(v)$ is sufficiently  high. Sooner or later, a
measurement gives the sufficiently big outcomes. The quantum
monitoring takes place. At this moment, the state of the qubit is
known with the accuracy desired and it is given by  the values of
outcomes $\hat{\rho}=\hat{\sigma_i}v_i/v$. Since
$\hat{\rho}^2=\hat{1}$, this is a pure state.

We stress that the monitoring does not constitute a single-shot
measurement of all components of the {\it initial} unknown quantum state.
This would be forbidden by the basic laws of quantum mechanics.
Indeed, the time required for an accurate monitoring exceeds by far
the measurement time of the detectors. By this time, the
initial state is completely forgotten.
However, the monitoring gives the observer complete information
about the final quantum state.

One also could argue that any pure state of the qubit can be
obtained in a simpler fashion. One would just choose a proper $H_q$
and wait for dissipation to bring the qubit to the state of lowest
energy. One could also try a projective measurement in a certain
basis: after several tries, such measurement would give the state
desired. We note however that for both approaches the resulting pure
state is a priori known to the observer. This is not the case of
quantum monitoring: here we let the quantum system to make "its own
choice" of the final pure state and do not enforce this choice by
any means.

\begin{figure}[htb]
  \begin{tabular}{cc}
   \resizebox{84mm}{!}{\includegraphics{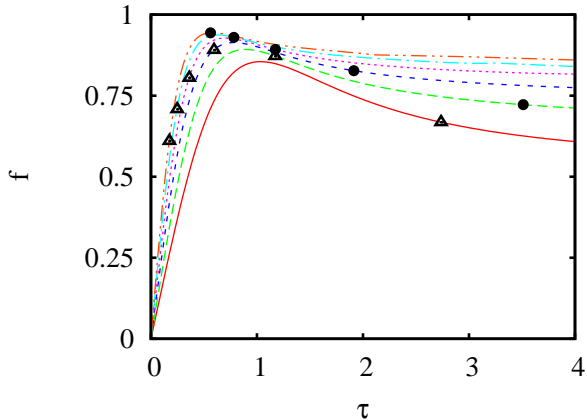}}
   \end{tabular}
\caption{(color online) Fidelity of quantum monitoring $f$ vs. the
measurement duration $\tau$  for quantum-limited detector. From
upper to lower curve $v$ are respectively  $4$, $3.5$, $3$, $2.5$,
$2$, and $1.5$.  Bullets  at each curve indicate the value of $\tau$
at which the probability to get the outcome larger than the
corresponding $v$ for each curve is $10\%$; triangles at each curve
indicate the value of $\tau$ at which the probability to get the
outcome larger than the corresponding $v$ for each curve is $50\%$.}
\end{figure}

The better the accuracy desired $a_{des}\equiv 1-f \ll 1$, the
bigger outputs are required, $v > C_d/a_{des}$. The typical waiting
time grows exponentially. To estimate it,  we assume  the duration
of each measurement $\tau$ is in the order of $\tau_m$.
This is because the longer durations are not favorable due
to decoherence.  The waiting time is inversely  proportional to the
probability to have sufficiently high outcomes $\tau_w \sim
\tau/P(v;\tau)$.
 Since the success probability  is exponentially small:
 $P(v;\tau)\sim \exp(-v^2 \tau/2)$
from the saddle point solution, we then estimate $\tau_w  \sim
\exp(v^2)$.  Therefore, we shall expect
$$
\mathrm{log}(\tau_w) \sim
a_{des}^{-2}.
$$

This estimation of
the waiting time of successful quantum monitoring
sounds pessimistic or at least causes a
doubt concerning the practical feasibility of the monitoring.
To prove that the monitoring is practical, we are going to show that a
reasonably high fidelity can be achieved in a reasonably short time.

The above arguments are based on the analytical
saddle point solution valid at
large $\tau$.
Now we investigate the measurements of moderate duration,
$\tau\sim 1 $. This can only be done by numerical calculation.
Solving Eqs.
(\ref{rho-chi-1}) and (\ref{rho-chi-2}), and making the Fourier
transformation according to Eq. (\ref{Fourier}),  we  evaluate and plot
$f(v)$ of a single measurement of duration $\tau$ versus $\tau$ for
the quantum-limited detector and the worse detector ($C_d=2$) in
Fig. 4 and Fig. 5
 respectively.
Each curve gives $f(\tau)$ at a given outcome $v$. At each curve, we
put a "bullet"  to indicate the duration of measurement $\tau$  at
which the probability  to obtain the outcomes larger than the given
one  is $10\%$.  Similarly, we put a "triangle" to indicate  $\tau$
at which the probability is $50\%$. To find the positions of the
symbols, for each $v$ we solve numerically for the values of $\tau$
that satisfy
\begin{eqnarray}
\frac{\int_{v}^{+\infty}\mathrm{Tr}\hat{\rho}(v',\tau) v'^2 d
v'}{\int_{0}^{+\infty}\mathrm{Tr}\hat{\rho}(v',\tau) v'^2 d v'}
=10\%\, (50 \%)
\end{eqnarray}
respectively.
 We see from the plots that $f=0.95$ is achieved for a
quantum-limited detector at $v=4$ and $\tau=0.7$. At these
parameters, $10\%$ of the measurements are successful, i.e. give the
output $v >4$. To achieve a success, the measurement
has to be repeated typically 10 times given the
success probability of $10\%$.
We conclude that the $5\%$ accuracy is typically
achieved in a time interval
$\simeq 10 \times 0.7 \tau_m= 7\tau_m$. This is not
much slower than the QND measurement of the same accuracy.

Let us explain the relation between the monitoring discussed
and the
quantum algorithms that use the results of partial measurement. These algorithms
have been introduced in the context of two-qubit systems
 \cite{Bennett,
Deutsch} but also may be applied to a general quantum state
\cite{Nakazato}. Speaking very generally, such algorithms start with
a quantum state, pure or mixed, and aim at producing another state
(pure and/or highly entangled). They proceed in steps. Each step
involves interaction with ancilla qubits that results in an
entanglement of the qubit and the ancillae. Importantly, the
projective measurement of ancilla qubit(s) is performed at each
step. The result of this measurement is used to {\it decide} upon
next step: possible decisions include to re-request the initial
state, to stop since the fidelity desired is reached, to apply a
certain quantum gate.

The quantum monitoring proceeds similarly. It starts with an almost
isotropic initial state: $\hat{\rho}=\hat{1}$.  The qubit is being
measured by our three independent quantum limit detectors during a
time interval $\tau$. The outcome of the detectors is used to {\it
make a decision}. If the sum of the square of the three detectors
output $v_i$ of  are small: $\sqrt{v_1^2+v_2^2+v_3^2}< 4$, the
measurement is disregarded: this is an analog of re-requesting the
initial state. The measurement is repeated until the values of the
outputs are sufficiently high $v=\sqrt{v_1^2+v_2^2+v_3^2}>4$. In
this case, the monitoring may stop since its goal is reached: the
qubit is purified to a state given by the detector outputs $v_{1}$,
$v_{2}$ and $v_{3}$.  Since the $v_i$ ($i=1,2,3$) are random, the
purified state of the qubit is also random.  Thus one could say that
the CWLM monitoring implements a quantum algorithm of the sort
described, in the same sense as an analogous computer may implement
discrete computer algorithms.

\begin{figure}[htb]
  \begin{tabular}{cc}
   \resizebox{84mm}{!}{\includegraphics{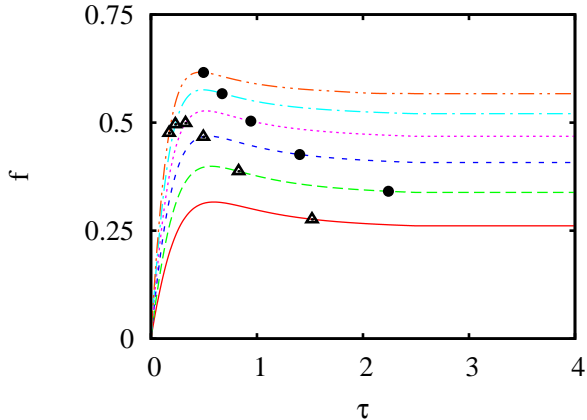}}
   \end{tabular}
\caption{(color online) Fidelity of quantum monitoring $f$ versus
the measurement duration $\tau$ for worse detector ($C_d=2$).
 From upper to lower curve $v$ are respectively  $4$, $3.5$, $3$, $2.5$,
$2$, and $1.5$.  Bullets  at each curve indicate the value of $\tau$
at which the probability to get the outcome larger than the
corresponding $v$ for each curve is $10\%$; triangles at each curve
indicate the value of $\tau$ at which the probability to get the
outcome larger than the corresponding $v$ for each curve is $50\%$.}
\end{figure}
\section{Summary}
In conclusion,  we have shown how to evaluate the full statistics of
the outcomes of a CWLM on a qubit. We are also able to evaluate
joint probability distribution of the outcomes and the qubit
variables after the measurement. For a single detector, we have
illustrated the QND measurements and understood the recent proposal
of quantum un-demolition measurement. Most interesting results
concern the simultaneous CWLM of three non-commuting variables by
three detectors. Such "measurement" is obtrusive and typically
scrambles the initial qubit wavefunction. However, we have
demonstrated a high degree of correspondence between the wave
function {\it after} the measurement and the outcomes of the three
detectors. Therefore, such CWLM may be used for high-fidelity
quantum {\it monitoring} of the qubit. The monitoring in fact
amounts to a {\it purification} of the qubit state in a random
direction $\vec{v}= (v_1,v_2,v_3)/|v|$ at Bloch sphere, $v_{1-3}$
being random outputs of the detectors. We have drawn analogy with
quantum  algorithms that use the outputs of ancilla measurements to
decide on the purification degree reached and the quantum gates to
be applied.

The interpretation we give to the results is of course
not the only possible one. The communications with
several colleagues have convinced us that "interesting" and
"important" defy an unambiguous definition as far as theory of
quantum measurement is concerned. In any case, we have developed
calculational tools to access the joint probability
of the qubit degrees of freedom and the outcomes of linear
detectors measuring the
qubit. We have also derived the representative results.
It is up to the reader to conclude.

\acknowledgements
H.W. acknowledges the financial support  in the framework
of NanoNed initiative (project DSC.7023).
Y.N. appreciates the participation in 2006 Aspen Summer
Program where he got the impetus to this work.

\appendix

\section{Derivation of the detector action}
The derivation of the path-integral representation for a set of
variables linear in boson creation/annihilation is a straightforward
task. It is instrumental in dissipative quantum mechanics and
therefore is to be found in basic literature on the subject.
References \onlinecite{Caldeira} are usually cited in this respect.
Owing to the simplicity of the problem, there are many other
derivations of the kind that are tailored to specific models (e.g.
\cite{Kleinert}) and usually assume thermal equilibrium of the boson
bath. To avoid any confusion, we present this part of the derivation
here. We do the derivation in the most general terms possible and
specify to the concrete model in use at the later stage of the
calculation.

Let ${X}_j$ be a set of the variables linear in boson
creation/annihilation operators, $\hat{X}_j(t)$ being the Heisenberg
time-dependent operators of these variables. Let us first disregard
the coupling with the qubit, so that the time dependence of
$\hat{X}_j(t)$ is governed by the detector Hamiltonian $H_d$ only.
Explicitly, the Heisenberg equation reads
$$
\frac{d \hat{X}_j(t)}{dt} = \frac{i}{\hbar} [H_d,\hat{X}_j(t)].
$$

We are interested in the generating function of the variables which
we present in the following form (c. f. Eq. \ref{action})
$$
Z(\{\chi_j(t)\}) = \mathrm{Tr}\bigl( \overrightarrow{\mathrm{T}} e^{
\frac{i}{2} \int dt\hat{X}_j(t)\chi_j(t)} \hat{\mathrm{\rho}_d}(0)
\,\overleftarrow{\mathrm{T}}e^{ \frac{i}{2}\int dt
\hat{X}_j(t)\chi_j(t)} \bigr) \,.
$$
We assume summation over repeating indexes $j$ and skip time indexes
of $\hat{X}(t)$, $\chi(t)$ for brevity. Differentiating with respect
to the parameters $\chi^{\pm}_j(t)$ of the generating function, one
reproduces all possible products of the operators $\hat{X}_j(t)$.

Let us introduce a path integral over the variables $X^{\pm}_j(t)$
associated with the operators $\hat{X}_j(t)$ by means of the
following identity:
\begin{eqnarray}
Z(\{\chi_j(t)\}) = \int \mathcal{D}{\bar X}^{+} \mathcal{D} {\bar
X}^{-} {\rm Tr} \left[ \overrightarrow{\mathrm{T}}
\prod\limits_{t,j} \delta(X_j^{-}(t) -\hat{X}_j(t))
e^{\frac{i}{2}\int dt ({X}^{-}_j + {X}^{+}_j) \chi_j}
\hat{\mathrm{\rho}_d}(0) \right.
\nonumber\\
\left. \overleftarrow{\mathrm{T}} \prod\limits_{t,j}
\delta(X_j^{+}(t) -\hat{X}_j(t)) \right]\,.
\label{delta-representation}
\end{eqnarray}
Here we insert $\delta$-functions that replace the operators
$\hat{X}_j(t)$ by the fluctuating fields $X_j^{\pm}(t)$, separately
for two parts of the Keldysh contour.

The use of this representation is that we can treat the coupling
between the detector and an arbitrary quantum system in the form of
the influence functional. If the coupling between the detector and
the system has the form
$$
H_{int} = \sum_j \hat{X}_j(t)\hat{S}_j;
$$
$\hat{S}_j$ being operators defined in the subspace of the quantum
system, we can formally substitute $\hat{X}_j(t) \to
{X}^{\pm}_j(t)$. The resulting influence functional thus reads
\begin{equation}
{\cal Z}_{Iq}( \{X^{-}_j(t)\},\{X^{+}_j(t)\}) = \mathop{{\rm
Tr}}\limits_S \bigl( \overrightarrow{\mathrm{T}} e^{
-\frac{i}{\hbar} \int dt{X}^-_j\hat{S}_j} \hat{\mathrm{\rho}_S}(0)
\,\overleftarrow{\mathrm{T}}e^{ \frac{i}{\hbar}\int dt
{X}^+_j\hat{S}_j} \bigr) \,,
\end{equation}
where the trace is over the subspace of the quantum system,
$\hat{\mathrm{\rho}_S}(0)$ is its initial density matrix and the
time dependence of $\hat{S}_j$ is governed by the separate
Hamiltonian on the subspace of the quantum system.

Let us turn to the evaluation of the representation
(\ref{delta-representation}). To facilitate the operations with
$\delta$-functions, we represent them by means of extra integration
over the auxiliary variables $k^{\pm}(t)/\hbar$. At each time
moment,
$$
\delta(X_j^\pm -\hat{X}_j) = \int \frac{dk_j^{\pm}}{2\pi}
e^{ik_j^+(X_j^\pm -\hat{X}_j)}.
$$
With these extra variables, the integral becomes
\begin{eqnarray}
Z(\{\chi_j(t)\}) = \int \mathcal{D}{\bar X}^{+} \mathcal{D} {\bar
X}^{-} \mathcal{D}{\bar k}^{+} \mathcal{D} {\bar k}^{-} e^{ {i}\int
dt{X}^{-}_j(k^{-}_j+\chi_j/2)} e^{ {i}\int
dt{X}^{+}_j(k^{+}_j+\chi_j/2)}
\nonumber\\
{\rm Tr}\left[ \overrightarrow{\mathrm{T}} e^{ -{i}\int    dt
{\hat{X}}_jk_j^{-}} \hat{\mathrm{\rho}_d}(0)
\overleftarrow{\mathrm{T}}
 e^{ -{i}\int dt {\hat{X}}_jk_j^{+}}\right].
\end{eqnarray}

Let us now take the trace over the boson degrees of freedom. To do
this, we use the widely known relation
$$
\langle e^{\hat{A}} e^{\hat{B}}\rangle =  e^{\langle
\frac{{\hat{A}^2 +\hat{B}^2}}{2} +\hat{A}{\hat{B}}\rangle}
$$
that holds for $\hat{A},\hat{B}$ that are linear in boson operators
under condition of Wick's theorem. This allows us to express the
trace in terms of the two-point correlators of boson variables,
$\langle \hat{X}_i(t) \hat{X}_j(t')\rangle$. We may assume $\langle
\hat{X}_j\rangle\equiv0$ without compromising generality. The
resulting expression reads
\begin{eqnarray}
Z(\{\chi_j(t)\}) = \int \mathcal{D}{\bar X}^{+} \mathcal{D} {\bar
X}^{-} \mathcal{D}{\bar k}^{+} \mathcal{D} {\bar k}^{-} e^{ {i}\int
dt{X}^{-}_j(k^{-}_j+\chi_j/2)}
e^{ {i}\int dt{X}^{+}_j(k^{+}_j+\chi_j/2)} e^{\tilde{A}_d};\nonumber \\
\tilde{A}_d = -\int dt dt'\left[ \frac{1}{2} k^{+}_i(t) k^{+}_j(t')
\langle\overleftarrow{\mathrm{T}}(\hat{X}_i(t) \hat{X}_j(t'))\rangle
+\frac{1}{2} k^{-}_i(t) k^{-}_j(t')
\langle\overrightarrow{\mathrm{T}}(\hat{X}_i(t) \hat{X}_j(t'))
\rangle \right.
\nonumber \\
\left. +k^{-}_i(t) k^{+}_j(t') \langle\hat{X}_i(t)\hat{X}_j(t')
\rangle\right].
\end{eqnarray}
It is instructive to introduce at this stage "classical" and
"quantum" variables defining $X^{\pm}_j=X_j \pm \hbar x_j/2$,
$k^{\pm}_j=K_j \pm k_j/2\hbar$. In these variables, the two-point
correlators are naturally collected to symmetrized noises
$S_{ij}(t,t')$ and Kubo-like response functions $A_{ij}(t,t')$,
$$
S_{ij}(t,t') = \frac{1}{2} \langle  \hat{X}_i(t)\hat{X}_j(t') +
\hat{X}_j(t')\hat{X}_i(t)\rangle;\; A_{ij}(t,t') =  -\frac{i}{\hbar}
\Theta(t-t') \langle [ \hat{X}_i(t),\hat{X}_j(t')] \rangle.
$$
and the action becomes
$$
\tilde{A}_d = -\int dt dt'\left[ 2 K_i(t)K_j(t') S_{ij}(t,t') + {i}
k_i(t)K_j(t') A_{ij}(t,t') \right].
$$
and does not contain $\hbar$. One can now integrate over the
auxiliary variables $K_j(t),k_j(t)$ to get the action in terms of
$X_j(t),x_j(t)$. This Gaussian integral can be readily taken by the
saddle-point method. For our model, it is convenient to make the
time-local approximation first. We replace the kernels $S_{ij},
A_{ij}$ with their time-local expressions Eqs.(\ref{responses},
\ref{noises}) and perform integration over $K,k$ to arrive at Eq.
\ref{newaction1}.

\section{Augmented Bloch-Redfield equation
and Lindblad form} In this appendix we cast the Eq.
(\ref{master-eq}) to the Lindblad form\cite{Lindblad} to illustrate
the entanglement of the qubit and detector and to prove the
positivity of the density matrix encompassing the qubit and output
variable of the detector.

To specify this underlying density matrix, it is useful to introduce
a quantum variable $\hat{p}$ defined as
$$
\hat{p}=\int_{0}^{t}\hat{V}(t')\, dt'\,.
$$
This variable represents the {\em integrated} detector output over
the interval $(0,t)$. It differs only by a factor $t$ from the
outcome $V_o$ of the measurement at the same time interval $(0,t)$
as we have defined in the main text.

We denote by $p$ and $|p\rangle$  the eigenvalue and eigenvector of
$\hat{p}$ respectively:  $\hat{p}|p\rangle=p|p\rangle$.  The subject
of our interest is the time-dependent reduced density matrix in the
space $|\mathrm{spin}\rangle\bigotimes |p\rangle $,
$\hat{\rho}(p,p';t)$, where "hat" denotes the matrix in the
pseudo-spin space. Since $p$ is related to $V_o$ upon a factor, this
matrix is equivalent to $\hat{R}(V_o,V'_o)$ used in the main text.
The quantity $\hat{\rho}(\chi)$ in the augmented Bloch-Redfield
equation is related to the diagonal part of this density matrix by
Fourier transform
\begin{eqnarray}
\hat{\rho}(p,p;t)=\int d\chi e^{-ip\chi}\hat{\rho}(\chi,t)\,.
\label{rho-p}
\end{eqnarray}
Making the inverse Fourier transform, we deduce from Eqs.
(\ref{master-eq}) and (\ref{rho-p}) the equation for
$\hat{\rho}(p,p;t)$:
\begin{eqnarray}
      \frac{\partial{\hat{\rho}}(p,p;t)}{\partial t} &=&
      -\frac{i}{\hbar} [H_q,\hat{\rho}(p,p;t)
      ]+\frac{S_{22}}{2}\frac{ \partial^2 \hat{\rho}(p,p;t)}{\partial
      p^2}\nonumber \\
& &-\frac{a_{21}}{2}\bigl(\frac{\partial \hat{\rho}(p,p;t)}{
\partial p} \hat{\sigma}_3 + \hat{\sigma}_3
\frac{\partial \hat{\rho}(p,p;t)}{ \partial p}\bigr) \nonumber \\
& &-\frac{iS_{12}}{\hbar}\bigl(\frac{\partial \hat{\rho}(p,p;t)}{
\partial p} \hat{\sigma}_3
         - \hat{\sigma}_3 \frac{\partial \hat{\rho}(p,p;t)}{\partial p}\bigr)\nonumber \\
& &-\frac{S_{11}}{\hbar^2}\bigl(\hat{\rho}(p,p;t)-  \hat{\sigma}_3
        \hat{\rho}(p,p;t)\hat{\sigma}_3
        \bigr) \,.
        \label{physics-eq}
\end{eqnarray}
This is an evolution equation in partial derivatives.

Next, we demonstrate
 that Eq. (\ref{physics-eq}) is indeed of
a Lindblad type. This also proves that the density matrix satisfying
the equation has positive diagonal elements. We work in the space
$|\mathrm{spin}\rangle\bigotimes |p\rangle$. We can check that the
operator $\hat{o}=i\frac{\partial}{\partial p}$  has the following
properties:
\begin{subequations}
\begin{eqnarray}
\langle p|\hat{\rho}\hat{o} |p'\rangle &=&-i \frac{\partial
\hat{\rho}(p,p';t) }{\partial
p'}\,, \label{id-1}\\
\langle p|\hat{o}\hat{\rho} |p'\rangle &=&i\frac{\partial
\hat{\rho}(p,p';t )}{\partial p}\,. \label{id-2}
\end{eqnarray}
\end{subequations}
The  Lindblad form of an evolution equation
reads\cite{Lindblad,Presilla}
\begin{eqnarray}
\frac{\partial \hat{\rho}}{\partial t}=-\frac{i}{\hbar}\bigl[H,
\hat{\rho}
\bigr]+\frac{1}{2}\sum_{\nu}\bigl([\hat{L}_{\nu}\hat{\rho},\hat{L}_{\nu}^{\dag}]+
[\hat{L}_\nu,\hat{\rho}\hat{L}_{\nu}^{\dag}] \bigr) \,,
\label{Lind-type}
\end{eqnarray}
where $H$ is a Hermitian operator, generally not coinciding with the
qubit Hamiltonian,  and $\hat{L}_{\nu}$ ($\nu=1,\cdots$) are
arbitrary operators. We need to prove that a proper choice of the
Lindblad operators $\hat{L}_{\nu}$ and the Hamiltonian $H$
reproduces Eq. \ref{physics-eq}.

To do so, we introduce $2$ Lindblad operators as follows:
\begin{subequations}
\begin{eqnarray}
\hat{L}_1&=&\sqrt{\frac{4S_{11}S_{22}-4S_{12}^2-\hbar^2
a_{21}^2}{4\hbar^2
S_{22} }}\,\hat{\sigma}_3 \,,\label{L1}\\
\hat{L}_2&=&\sqrt{S_{22}}\bigl(\hat{o}-\frac{S_{12}}{\hbar
S_{22}}\hat{\sigma}_3 -\frac{ia_{21}}{ 2S_{22}}\hat{\sigma}_3
\bigr)\,,\label{L2}
\end{eqnarray}
\end{subequations}
here $4S_{11}S_{22}-4S_{12}^2-\hbar^2 a_{21}^2\geq 0$ is guaranteed
by the Cauchy-Schwartz inequality (see Sec. II). The Hermitian
operator reads:
\begin{eqnarray}
H=H_{q}-\frac{\hbar a_{21}}{2}\hat{o}\hat{\sigma}_3\,.
\end{eqnarray}
where $H_q$ is the qubit Hamiltonian. We substitute the operators to
Eq. (\ref{Lind-type}) to obtain the following:

\begin{eqnarray}
\frac{\partial \hat{\rho}(p,p';t)}{\partial
t}&=&-\frac{i}{\hbar}\bigl[H,\hat{\rho}(p,p';t) \bigr]\nonumber\\
& &+\frac{S_{22}}{2}\bigl(2 \frac{\partial^2
\hat{\rho}(p,p';t)}{\partial p
\partial p'}+\frac{\partial^2 \hat{\rho}(p,p';t)}{\partial p^2}
 +\frac{\partial^2 \hat{\rho}(p,p';t)}{\partial p'^2}\bigr)\nonumber\\
& &-\frac{a_{21}}{2}\bigl(\frac{\partial \hat{\rho}(p,p';t)
}{\partial p'}\hat{\sigma}_3+\frac{\partial \hat{\rho}(p,p';t)
}{\partial p}\hat{\sigma}_3 +\hat{\sigma}_3 \frac{\partial
\hat{\rho}(p,p';t) }{\partial p}+\hat{\sigma}_3 \frac{\partial
\hat{\rho}(p,p';t) }{\partial
p'} \bigr)\nonumber \\
& & -\frac{iS_{12}}{\hbar}\bigl(\frac{\partial
\hat{\rho}(p,p';t)}{\partial p}\hat{\sigma}_3+\frac{\partial
\hat{\rho}(p,p';t)}{\partial
p'}\hat{\sigma}_3-\hat{\sigma}_3\frac{\partial
\hat{\rho}(p,p';t)}{\partial p}- \hat{\sigma}_3\frac{\partial
\hat{\rho}(p,p';t)}{\partial p'}
  \bigr)\nonumber \\
  & &+\frac{S_{11}}{\hbar^2}\bigl( \hat{\sigma}_3
\hat{\rho}(p,p';t)\hat{\sigma}_3-\hat{\rho}(p,p';t) \bigr) \,.
  \label{lin-matrix}
\end{eqnarray}

We made use of the properties (\ref{id-1}) and  (\ref{id-2}). We
have not done yet, since the above equation is for a two-indexed
density matrix $\hat{\rho}(p,p')$ and not for its diagonal part. We
still have to prove that the above equation does not mix the
diagonal and non-diagonal elements. So that, to relate Eq.
(\ref{physics-eq}) with Eq. (\ref{lin-matrix}), we change variables
as follows:
\begin{subequations}
\begin{eqnarray}
P_s&=&\frac{p+p'}{2}\,, \\
P_d&=&\frac{p-p'}{2}\,.
\end{eqnarray}
\end{subequations}
Thus,
\begin{subequations}
\begin{eqnarray}
\frac{\partial }{\partial p}&=&\frac{1}{2}(\frac{\partial}{\partial
P_s}+\frac{\partial}{\partial P_d}) \,,\label{pp1}
  \\
\frac{\partial}{\partial p'}&=&\frac{1}{2}(\frac{\partial}{\partial
P_s}-\frac{\partial}{\partial P_d}) \,. \label{pp2}
\end{eqnarray}
\end{subequations}
From Eqs. (\ref{lin-matrix}), (\ref{pp1}) and (\ref{pp2}) we obtain:
\begin{eqnarray}
\frac{\partial \hat{\rho}(P_s,P_d;t)}{\partial
t}&=&-\frac{i}{\hbar}\bigl[H,\hat{\rho}(P_s,P_d;t)
\bigr]+\frac{S_{22}}{2} \frac{\partial^2
\hat{\rho}(P_s,P_d;t)}{\partial P_s^2} \nonumber \\
 & &-\frac{a_{21}}{2}\bigl(\frac{\partial \hat{\rho}(P_s,P_d;t)
}{\partial P_s}\hat{\sigma}_3+\hat{\sigma}_3 \frac{\partial
\hat{\rho}(P_s,P_d;t) }{\partial
P_s} \bigr)\nonumber \\
& & -\frac{iS_{12}}{\hbar}(\frac{\partial
\hat{\rho}(P_s,P_d;t)}{\partial
P_s}\hat{\sigma}_3-\hat{\sigma}_3\frac{\partial
\hat{\rho}(P_s,P_d;t)}{\partial P_s})  \nonumber \\
& &+\frac{S_{11}}{\hbar^2} \hat{\sigma}_3
\hat{\rho}(P_s,P_d;t)\hat{\sigma}_3\,.\label{lindblad-eq}
\end{eqnarray}
Eqs. (\ref{physics-eq}) and (\ref{lindblad-eq}) has the same form.
Therefore, we have proved that $\hat{\rho}(p,p;t)$ satisfies the
Lindblad equation in the  $|\mathrm{spin}\rangle\bigotimes |p\rangle
$ space matrix form and thus is positive.

We can straightforwardly extend the above scheme to Eq.
(\ref{master-eq2}) of Sec. II that is valid for  the case  of three
detectors. In this case, we introduce  $6$ Lindblad  operators
 \begin{subequations}
\begin{eqnarray}
\hat{L}_{1i}&=&\sqrt{\frac{4S_{11}S_{22}-4S_{12}^2-\hbar^2
a_{21}^2}{4\hbar^2
S_{22} }}\,\hat{\sigma}_i\,, \label{3L1}\\
\hat{L}_{2i}&=&\sqrt{S_{22}}\bigl(\hat{o}_i-\frac{S_{12}}{\hbar
S_{22}}\hat{\sigma}_i -\frac{ia_{21}}{
2S_{22}}\hat{\sigma}_i\bigr)\,,\label{3L2}
\end{eqnarray}
\end{subequations}
where $i=1,2,3$. The Hermitian operator reads:
\begin{eqnarray}
H=H_{q}-\frac{\hbar a_{21}}{2}\sum_{i=1}^{3}\hat{o}_i\,,
\end{eqnarray}
where $\hat{o}_{i}=i\frac{\partial}{\partial p_i}$ and $p_i$ are the
eigenvalues of the operator $\hat{p}_i=\int_{0}^{t}\hat{V}_i(t')\,
dt'$ ($i=1,2,3$) corresponding to each detector.

To conclude, we have shown that the density matrix
$\hat{\rho}(p,p';t)$ in the space $|\mathrm{spin}\rangle\bigotimes
|p\rangle $ satisfies the Lindblad equation and thus its diagonal
part is positive. The diagonal part of the matrix is related to the
quantity $\hat\rho(\chi,t)$ in the Sec. II by the Fourier transform.
We note that Lindblad operators Eqs. (\ref{L1}, \ref{L2}, \ref{3L1},
\ref{3L2}) include both the degree of qubit  and the degree of the
detector(s). This signals the entanglement of the qubit and
detector(s)  In this sense, we measure both the qubit and
detector(s) in the context of FCS theory.

\end{document}